# SPLAY-BEND ELASTIC INEQUALITIES SHAPE TACTOIDS, TOROIDS, UMBILICS, AND CONIC SECTION WALLS IN PARAELECTRIC, TWIST-BEND, AND FERROELECTRIC NEMATICS*

Oleg D. Lavrentovich,

Advanced Materials and Liquid Crystal Institute,

Materials Science Graduate Program,

Department of Physics,

Kent State University, Kent, OH 4424

**Abstract.** Elastic constants of splay $K_{11}$, twist $K_{22}$, and bend $K_{33}$ of nematic liquid crystals are often assumed to be equal to each other in order to simplify the theoretical description of complex director fields. Here we present examples of how the disparity of $K_{11}$ and $K_{33}$ produces effects that cannot be described in a one-constant approximation. In a lyotropic chromonic liquid crystal, nematic droplets coexisting with the isotropic phase change their shape from a simply-connected tactoid to a topologically distinct toroid as a result of temperature or concentration variation. The transformation is caused by the increase of the splay-to-bend ratio $K_{11}/K_{33}$. A phase transition from a conventional nematic to a twist-bend nematic implies that the ratio $K_{11}/K_{33}$ changes from very large to very small. As a result, the defects caused by an externally applied electric field change the deformation mode of optic axis from bend to splay. In the paraelectric-ferroelectric nematic transition, one finds an inverse situation: $K_{11}/K_{33}$ changes from small to large, which shapes the domain walls in the spontaneous electric polarization field as conic sections. The polarization field tends to be solenoidal, or divergence-free, a behavior complementary to irrotational curl-free director textures of a smectic A.

# INTRODUCTION

Orientational ordering of liquid crystals brings about an important concept of Frank elastic constants $K_{ii}$ describing the energy cost of the gradients in molecular orientations. The Frank constants are of the dimension of a force, and thus can be represented as the ratio of some energy $U$ to a characteristic length $l$. In a conventional uniaxial nematic (N$_U$) formed by rod-like molecules, the latter can only be the molecular length, $l \sim 1$ nm. The energy $U$, as suggested by P.G. de Gennes [1], should be on the order of $k_B T_c$, where $k_B$ is the Boltzmann constant and $T_c$ is the clearing temperature at which the nematic transitions into an isotropic fluid. For $T_c \sim 300$ K, one finds $K_{ii} \sim \frac{U}{l} \sim 4$ pN, which is close to the experimentally measured values. For example, elastic constants of pentylcyanobiphenyl (5CB) are listed [2] at 305 K (about 4 K below the clearing point) as $K_{11} = 4.5$ pN for splay, $K_{22} = 3$ pN for twist, and $K_{33} = 5.5$ pN for bend. In 5CB, as in many other nematics formed by rod-like molecules, the constants follow the trend $K_{33} > K_{11} > K_{22}$ [3, 4, 5, 6]. A relevant geometrical parameter is the aspect length/diameter ratio, which justifies $\frac{K_{11,22}}{K_{33}} < 1$ [3, 4, 5]. $K_{22}$ is somewhat smaller than the other two moduli, which explains why twist often replaces splay and bend in nematic samples deformed by confinement, such as droplets of thermotropic [7, 8, 9, 10, 11] and lyotropic nematics [12, 13]. Although the occurrence of twist in chemically achiral materials [14, 15] is a very interesting topic awaiting its further exploration in the newly discovered ferroelectric nematics [16, 17], this review limits itself to recently described effects caused by a disparity of the splay $K_{11}$ and bend $K_{33}$ moduli. The Frank-Oseen free energy density corresponding to different bulk modes of distortions writes

$$f = \frac{1}{2}K_{11}(\text{div } \hat{\mathbf{n}})^2 + \frac{1}{2}K_{22}(\hat{\mathbf{n}} \cdot \text{curl } \hat{\mathbf{n}})^2 + \frac{1}{2}K_{33}(\hat{\mathbf{n}} \times \text{curl } \hat{\mathbf{n}})^2. \tag{1}$$

The splay and bend geometries are schematized in Fig.1.

There are numerous reasons why $K_{11}$ and $K_{33}$ can be different. If one assumes that the molecules are rigid, then the increase of the length/diameter aspect ratio might decrease $K_{11}/K_{33}$. However, as noted by R.B. Meyer [18], nematics formed by very long polymer chains should exhibit $K_{11} \gg K_{33}$ since splay creates empty spaces, which must be filled by the ends of molecules to keep the material's density constant. If the molecules are banana-like in shape, a similar inequality, $K_{11} > K_{33}$, is found experimentally [19, 20, 21, 22, 23, 24, 25, 26, 27] and theoretically [4, 28, 29, 30]; see also the recent reviews [31, 32]. An opposite and even stronger disparity of elastic constants is observed in an N$_U$ formed by acute-angle bent core molecules of a shape resembling a letter λ [33]. The measured splay constant is anomalously weak, $K_{11} = 2$ pN, significantly smaller than the bend constant $K_{33} = 15$ pN and even the twist constant $K_{22} = 5$ pN. The smallness of $K_{11}$ leads to a pronounced bias of defects towards configurations with splay [32, 33].

The low energy cost for bend in banana-like molecules inspired R.B. Meyer [18], I. Dozov [34], R. Memmer [35], and S. M. Shamid et al. [36] to predict the so-called twist-bend nematic ($N_{TB}$), experimentally found in materials formed by flexible dimeric [29, 37, 38] and rigid bent-core [39] molecules. Another notable result emerging from the smallness of $K_{33}$ is the formation of an oblique helicoidal cholesteric in an external electric [40, 41] or magnetic field [42], predicted by R.B. Meyer [43] and P.G. de Gennes [44].

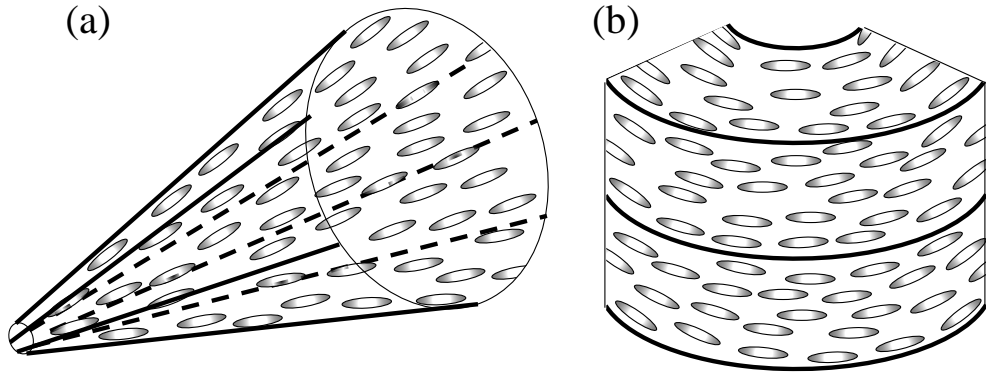

**Fig.1.** Director field in (a) splay and (b) bend.

In what follows, the presentation discusses (1) topological transformation of a lyotropic chromonic nematic droplet from a tactoid into a toroid driven by the increase of $K_{11}/K_{33}$ changes from large to small, (2) the geometry of defects formed in response to an electric field , which is controlled by a change from $\frac{K_{11}}{K_{33}} \gg 1$ to $\frac{K_{11}}{K_{33}} \ll 1$ as a result of the nematic-to-twist-bend nematic phase transition, and (3) conic sections such as parabola and hyperbola, which represent domain walls in thin films of ferroelectric nematics ($N_F$) shaped by $\frac{K_{11}}{K_{33}} \gg 1$.

# 1. Tactoid to toroid reshaping of a nematic droplet: From $K_{11} \approx K_{33}$ to $K_{11} > K_{33}$.

A lyotropic chromonic liquid crystal (LCLC) represents a dispersion of disk-like organic molecules in water. Hydrophobic cores of the molecules stack on top of each other, forming elongated aggregates which align parallel to each other [45]. Elastic constants in the $N_U$ phase depend strongly on the concentration and temperature [2, 46, 47]. One notable experimental finding is that the twist constant $K_{22}$ in the chromonic $N_U$ is anomalously low, less than 1 pN [2, 46, 47]; similar trend is established also for two other lyotropic systems, namely, solutions of the polymer poly-gamma-benzyl-glutamate [48] and the $N_U$ formed by disk-like micelles of surfactant molecules [49]. The strong dependence of $K_{11}$ and $K_{33}$ on the temperature and concentration can be interpreted in terms of the varying contour length of the aggregates $L$ and their persistence length $\lambda$, i.e., bending flexibility, $K_{11}/K_{33} \propto L/\lambda$ [2, 46]. The increase of $K_{11}$ with $L$ is expected on the grounds of the R.B. Meyer's argument that splay of long molecules is difficult from the entropy point of view, as it creates vacancies that should be filled with the ends of the molecules (or aggregates in the case of LCLCs) [18, 50]. Bend of long rigid rods might create similar problems, but these could be avoided if the molecules (aggregates) are easy to bend, i.e., when $\lambda$ is small. The anomalous smallness of $K_{22}$ can be qualitatively explained by the fact that twist does not create any "vacancies" if the aggregates arrange in layers perpendicular to the twist axis.

LCLCs exhibit broad biphasic regions in which the $N_U$ (or columnar) phase coexists with the isotropic phase. In coexistence, the aggregates are partitioned between the ordered and disordered phases, with longer aggregates residing in the condensed phase. Prior studies established that the increase of the chromonic concentration $c$ in a homogeneous $N_U$ phase of DSCG increases $K_{11}/K_{33}$ [46]. In the condensed $N_U$ droplets, an increase of the temperature results in a higher concentration of DSCG [51, 52], which in its turn, produces a larger $K_{11}/K_{33}$ responsible for the transformation of the $N_U$ droplets from a sphere-like tactoid to a torus-like toroid [12, 52], Fig.2. These two shapes are topologically distinct, as described by Euler characteristic $\chi$, calculated as $\chi = 2 - 2g$, where $g$ is the number of "handles"; a sphere has no handles, thus $\chi=2$, while a torus is a single handle, thus $\chi=0$.

The biphasic LCLC in Fig.2 represents a water dispersion of disodium cromoglycate (DSCG), of a concentration $c = 0.34$ mol/kg, with an added polyethylene glycol (PEG) as a condensing agent, at the concentration 0.012 mol/kg. The specimen is made deliberately thin so that the transformation of a thin disk-like tactoid into a torus with a well-defined and wide isotropic central region is clearly visible under a microscope, Fig.2. It starts with the detachment of the two surface point defects-boojums from the cusps of tactoid, making them two disclinations of strength +1/2 each. The disclinations approach each other and coalesce, forming a toroid with a large central isotropic region. A similar transformation is observed when the concentration of PEG increases [52].

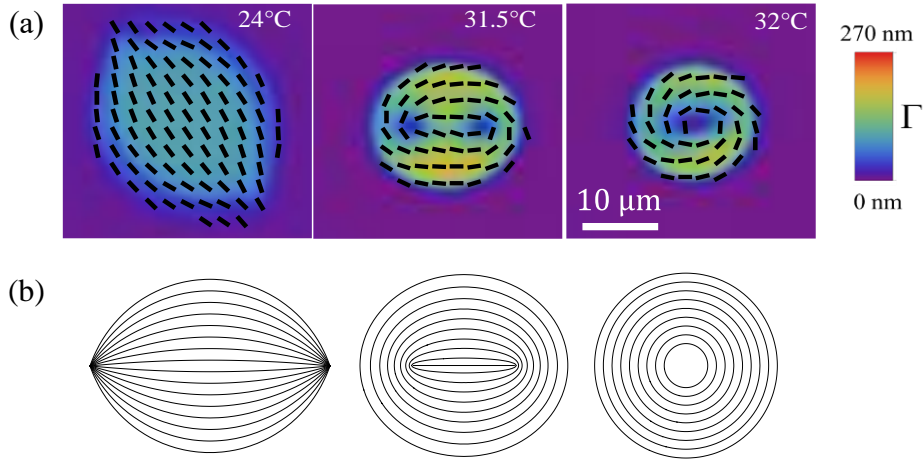

**Fig.2.** Temperature-triggered tactoid-to-toroid transformation: (a) experimental PolScope textures visualizing the director field; (b) director schemes for analytical estimates. Data from Ref. [52].

As follows from Eq.(1), the elastic energy of a toroid of a radius $a$ and a thickness $h$ with the director written in cylindrical coordinates as $\hat{\mathbf{n}} = \{n_r, n_\psi, n_z\} = \{0,1,0\}$, depends only on the elastic constant of bend, $F_{tor} = \pi h K_{33} \ln\left(\frac{a}{r_i}\right)$, where $r_i$ is the radius of the isotropic core. The energy of a circular tactoid of the same volume $hA = \pi h a^2$ and the same surface area $A$ depends also on $K_{11}$, since splay is present at the cusps, Fig.2b: $F_{tac} \approx \frac{\pi h}{2} K_{11} \ln \frac{2a}{r_{cb}} + \frac{\pi h}{2} K_{33}(1 - \ln 2)$, where $r_{cb}$ is the radius of the core of the boojums [52]. As $K_{11}/K_{33}$ increases, the first term in the energy $F_{tac}$ increases and the tactoid becomes less energetically favorable as compared to the toroid. The transition condition is $K_{11}/K_{33} > 2$ for $a = 15$ μm, $r_i = r_{cb} = 2$ μm. Surface tension can also contribute to the transformation scenario, but its effect is weaker than that of the elasticity [52]. Numerical simulations, which account for both the elastic and surface energy consideration, provide a more accurate description of the transition, including the appearance and coalescence of the ½-disclinations [52].

Out-of-equilibrium and living systems show topological transformations in which $\chi$ changes. A cell dividing into two increases the net Euler characteristic from $\chi=2$ to $\chi=4$. An inverse process, a reduction of $\chi$, in which holes are pierced into a sphere, is involved in morphogenesis of multicellular organisms that develop from a spherical cell into torus-like or more complicated multiply-connected bodies [53, 54]. The mechanisms by which living matter employs surface and bulk forces to change topology, especially by decreasing $\chi$, are far from being understood.

Liquid crystal droplets represent a simple model system in which the effect of the bulk and surface forces on the shapes and the internal structure is, in principle, tractable. Droplets of thermotropic liquid crystals dispersed in an immiscible isotropic fluid such as glycerin [55] or in a polymer matrix [56] exhibit a spheroidal shape, $\chi = 2$, imposed by a strong interfacial tension, with a complex interior pattern of molecular orientation that depends on the preferred alignment at the surface. Wei et al. [57] and Peddireddy et al. [58] report on the shape change of $N_U$ droplets from a sphere to branched filamentous networks as a result of a reduction of surface tension. This transformation preserves $\chi=2$. Liquid crystal droplets could also divide at phase transitions, thus increasing $\chi$ from 2 to 4, 6, etc., as demonstrated for cholesteric droplets during a transition to a smectic A phase [59]. The tactoid-to-toroid topological transformation [52] adds to this list. When $K_{11} \sim K_{33}$, the droplet accommodates both splay and bend of the director $\hat{\mathbf{n}}$ within a simply-connected tactoid; when $K_{11}$ increases, the droplet could afford only bend, which results in a torus-like shape with a hole in the center, Fig.2.

## 2. Nematic to twist-bend nematic: From $K_{11} \gg K_{33}$ to $K_{11} \ll K_{33}$.

The twist-bend nematic ($N_{TB}$) formed by flexible dimers or banana-like molecules exhibits a director field in the shape of a helicoid, maintaining a constant oblique angle $0 < \theta_0 < \pi/2$ with the helix axis $\hat{\chi}$, which we direct along the $z$ axis: $\hat{\mathbf{n}} = \{n_x, n_y, n_z\} = \{\sin\theta_0 \sin\varphi, \sin\theta_0 \cos\varphi, \cos\theta_0\}$, where $\varphi = t_{tb}z$ is the azimuthal angle, $t_{tb} = 2\pi/p_{tb}$, $p_{tb} \sim 10$ nm is the pitch of the helicoid. The reason for this structure is the tendency of molecules to induce a local bend [18, 34, 35, 36]. A pure bend of a constant curvature $|\hat{\mathbf{n}} \times \text{curl}\hat{\mathbf{n}}| = const$, however, cannot fill the space. Geometrically, bend $\hat{\mathbf{n}} \times \text{curl}\hat{\mathbf{n}}$ is a vector along the principal normal to the line that envelops the spatially-varying director [60]. The length of this vector at point M is the bend curvature of the line at that point. To maintain a constant bend in space, the line should be of a helicoidal shape. Such a line can be defined on a circular cylinder surface, directed at a constant angle to the axis [31]; this geometry implies twist, which enables a constant bend, hence the name of $N_{TB}$.

The $N_{TB}$ is typically observed upon cooling of an $N_U$; in the latter, the bend tendency manifests itself in a very small $K_{33}$, which makes $K_{11}/K_{33}$ as high as 30 [25], [31]. Once the $N_{TB}$ emerges upon cooling, the bend could exist only as a nanoscale deformation of the director $\hat{\mathbf{n}}$ but not as a macroscopic deformation of the helicoidal axis $\hat{\chi}$. The reason is that $\hat{\chi}$ is perpendicular to surfaces of a constant azimuthal angle $\varphi$: any bend or twist of $\hat{\chi}$ changes the equilibrium $p_{tb}$, i.e., violates the equidistance of the nanoscale $N_{TB}$ pseudolayers. As a result, the optic axis in the $N_U$, which is the director $\hat{\mathbf{n}}$, and the optic axis in the $N_{TB}$, which is $\hat{\chi}$, show dramatically different textures in response to confinement or to an external field, Fig.3.

In the $N_U$ of an $N_{TB}$-forming mesogen, such as DTC-C9 in Fig.3g, whenever there is a choice between splay and bend, the latter is realized. A good illustration is a Frederiks transition in a sandwich-type cell with homeotropic anchoring. If the material is of a negative dielectric anisotropy, $\Delta\varepsilon < 0$, an electric field applied along the normal to the cell causes bend distortions in the vertical plane, which appear as umbilics of a topological charge $\pm 1$, Fig.3a,e [38]. In the plane of the cell, the director around +1 defects show a clear preference for bend, Fig.3a, which is understandable since $K_{33} \ll K_{11}$. Once the material is cooled down to the $N_{TB}$ phase, the electric field-induced textures of $\hat{\chi}$ are very different, Fig. 3b,c,d,f. Above some threshold voltage, the $N_{TB}$ nucleates circular domains structurally similar to toric focal conic domains in a smectic A, Fig.3c. Further increase of the voltage transforms the circular domains into elongated "oily streaks" which expand and fill the space between the bounding plates [38, 61]. The textures of the optic axis $\hat{\chi}$ show a clear preference to splay, both in the plane of the cell, Fig.3b,c,d, and in the vertical cross-section, Fig.3f. In other words, the $N_U$-to-$N_{TB}$ transition is accompanied by a change from $\frac{K_{11}}{K_{33}} \gg 1$ for the director $\hat{\mathbf{n}}$ in the $N_U$ to $\frac{K_{11}}{K_{33}} \ll 1$ for the helicoidal axis $\hat{\chi}$ of the $N_{TB}$.

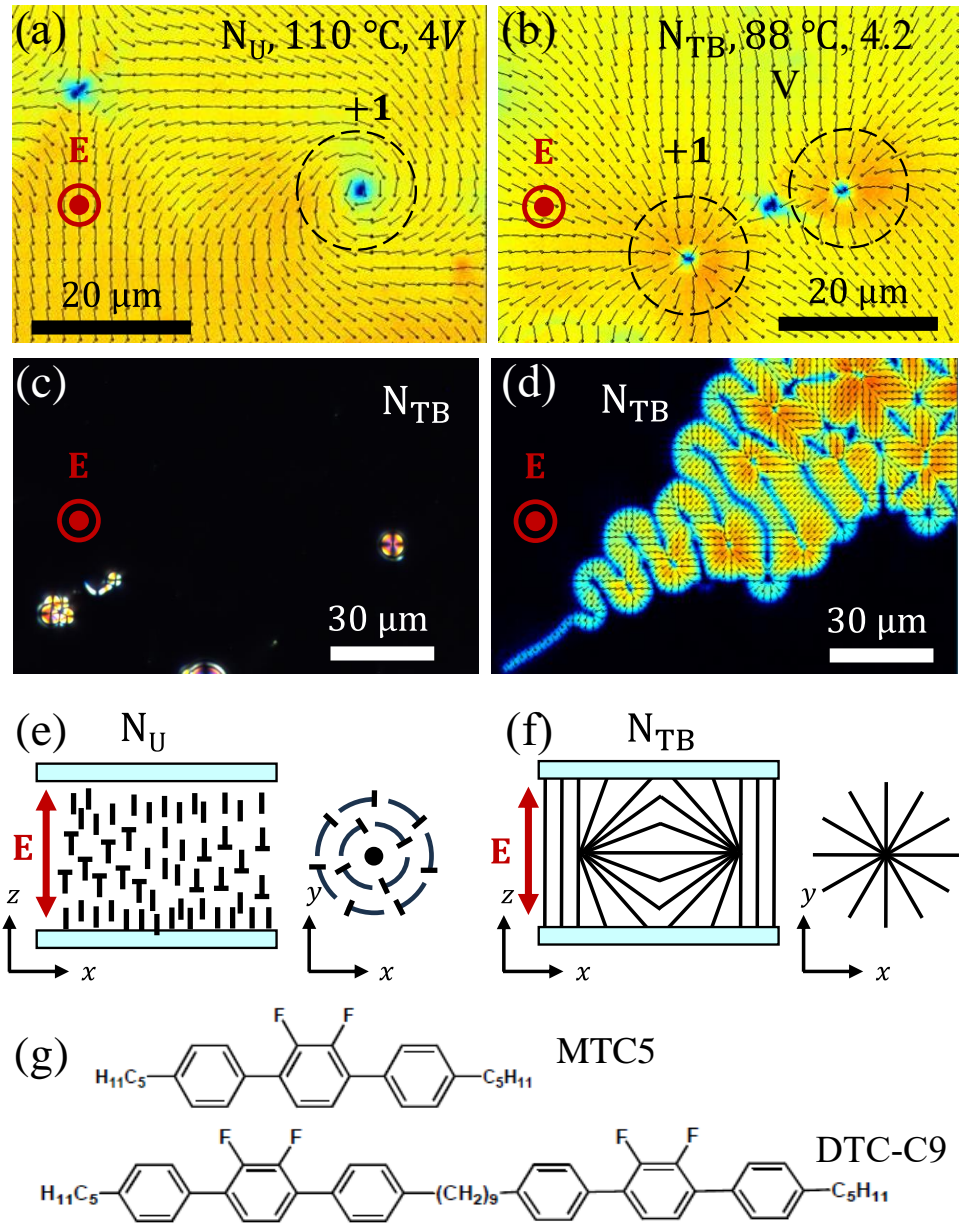

**Fig.3.** Electric field response of the $N_U$ and $N_{TB}$ with negative dielectric anisotropy in a homeotropic cell: (a) umbilics of the bend Frederiks transition in the $N_U$ caused by the electric field; note the predominance of bend around +1 defect; (b) a field-induced texture of the heliconical axis $\hat{\chi}$ in the $N_{TB}$; note the predominance of splay; (c) nucleation and (d) expansion of the realigned $N_{TB}$ structure with splay of $\hat{\chi}$; (e) a scheme of bend in an $N_U$ umbilic; (f) splay of $\hat{\chi}$ in the vertical cross-section of the cell; (g) chemical structure of the $N_{TB}$-forming composition DTC-C9:MTC5=7:3 by weight. Data from Ref. [38].

# 3. Nematic to ferroelectric nematic: From $K_{11} \ll K_{33}$ to $K_{11} \gg K_{33}$.

Pioneering exploration [62] of the nematic RM734 formed by molecules with a strong longitudinal dipole moments, ~10 D, revealed that the splay constant $K_{11}$ is very small in the N$_U$ phase of this material. The new phase that emerged from the N$_U$ upon cooling, later identified as the uniaxial ferroelectric nematic N$_F$ [63], exhibited textures of domains with oppositely directed spontaneous electric polarization **P** in planar cells.

In the N$_F$, the polarization vector is collinear with the director $\hat{\mathbf{n}}$. Splay is difficult since it produces a bound charge of density $\rho_b = -\text{div } \mathbf{P}$, which increases the electrostatic energy. As envisioned by R.B. Meyer [64] and detailed theoretically in the subsequent studies [65, 66, 67], $K_1$ associated with $(\text{div } \hat{\mathbf{n}})^2$ is renormalized for the distortions developing over length scales longer than the Debye screening length $\lambda_D$: $K_1 = K_{1,0}(1 + \lambda_D^2/\xi_P^2)$, where $K_{1,0}$ is the bare modulus, of the same order as the one normally measured in a conventional paraelectric N$_U$, $\xi_P = \sqrt{\frac{\varepsilon\varepsilon_0 K}{P^2}}$ is the so-called polarization penetration length, $\varepsilon\varepsilon_0$ is the dielectric permittivity of the material, $\varepsilon_0$ is the electric constant, $\lambda_D = \sqrt{\frac{\varepsilon\varepsilon_0 k_B T}{ne^2}}$, $e$ is the elementary electric charge, $n$ is the concentration of ions. For polarization density [68] $P = 5 \times 10^{-2} \text{C/m}^2$ and assumed $\varepsilon\varepsilon_0 \sim 10^{-9} \frac{C^2}{J \times m}$ and $K \sim 10^{-11}$ N, one finds a very short $\xi_P \approx 2$ nm. At the same time, the Debye screening length is expected to be larger: At $T=400$ K, and $n \leq 10^{23}/m^3$, $\lambda_D \geq 10$ nm. Therefore, the enhancement factor $\frac{\lambda_D^2}{\xi_P^2}$ could be strong and the ratio $K_1/K_3$ in N$_F$ could be significantly larger than 1.

Very little is known about the elastic constants in the N$_U$ phase of ferroelectric materials and practically nothing is known about the elasticity of N$_F$. Chen et al [69] measured $K_1 \approx 10 K_2$ in the N$_U$ phase of ferroelectric material DIO and expected [67] $K_1 \approx 2$ pN. Mertelj et. al. [62] reported that in the N$_U$ phase of RM734, $K_1$ is even lower, about 0.4 pN. Since the bend constant $K_3$ of N$_F$ is not expected to experience electrostatic renormalization, it could be a few tens of pN; Mertelj et. al. [62] found $K_3 \approx 10$-20 pN for the N$_U$ phase of RM734.

The role of space charge in shaping elastic anisotropy of bend vs splay has been extensively studied in the past for the ferroelectric chiral smectic C* (SmC*) [66, 70, 71, 72, 73]. Since the polarization vector in SmC* is perpendicular to the long axes of molecules, electrostatic effects lead to a large $K_3$, as discussed by Link et al. [71] and Pattanaporkratana et al. [72] for -1 disclinations, Zhuang [70] and Dolganov et al. [74] for $2\pi$ domain walls. In the $N_F$, electrostatic effects increase $K_1$ rather than $K_3$ since **P** is parallel to the long molecular axes; besides, these effects might be stronger than in the SmC* since the polarization of the $N_F$ is higher.

A qualitative evidence that $K_1 > K_3$ in the $N_F$ is presented by the textures of planar monocrystalline $N_F$ samples with air bubbles trapped between glass plates [75], Fig.4, domain walls in planar cells with rubbed substrates [76] and walls in thin azimuthally degenerate films in which the spatial variations of the polarization are not restricted by the externally imposed rubbing directions, Fig.5 [77].

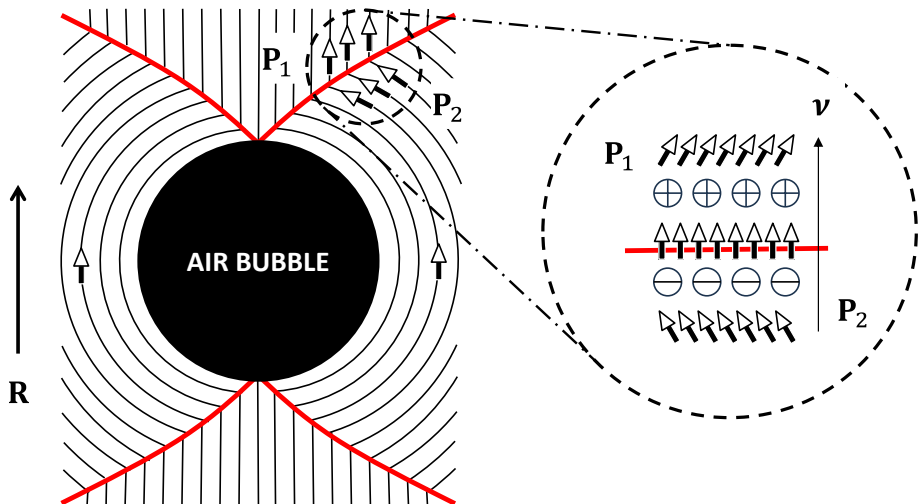

**Fig.4.** Parabolic domain walls caused by an air bubble in a planar $N_F$ cell. The parabola bisects the two polarization fields, a uniform $\mathbf{P}_1$ and a circular $\mathbf{P}_2$. The inset shows how the in-plane realignment of **P** leads to two sheets of opposite charges which attract each other thus stabilizing the wall. **R** is the rubbing direction on both plates of the sandwich-type sample; **ν** is an axis perpendicular to the wall. Redrawn from [75].

In a planar cell with a trapped air bubble, Fig.4, the polarization vector is subject to a frustration between the unidirectional rubbing of the substrates which aligns **P** uniformly along the rubbing direction **R**, and the circular interface of the bubble, which aligns **P** tangentially to itself. The frustration is resolved by four parabolic branches, emerging from the poles of the droplet and separating the domain of a circular $\mathbf{P}_2$ around the air bubble and the uniform far-field $\mathbf{P}_1$ dictated by the rubbing, Fig.4. This structure avoids splay thus the space charge is minimum, and the parabolic domain walls avoid being charged since they bisect the uniform and circular polarization fields [75]. In the description of the domain wall defect, one often assumes that the polarization vector remains in the plane of the wall [70, 72, 75], i.e., in the plane of Fig.4. If that is the case, then the projection $\mathbf{P} \cdot \mathbf{v} = P_v$ of the polarization vector onto the axis $\mathbf{v}$ normal to the wall increases while transitioning from $\mathbf{P}_1$ to $\mathbf{P}_2$. As a result, the derivative $-\partial P_v/\partial v$, which is the space charge density, produces two oppositely charged sheets attracting each other [70, 72, 75]. Electrostatic attraction is opposed by orientational elasticity [70, 72, 75]. The balance of the two yields an estimate of the domain wall width as the polarization penetration length [70], $\xi_P = \sqrt{\frac{\varepsilon\varepsilon_0 K}{P^2}} \sim 1$ nm, which is very short [70, 72, 75].

It turns out that parabolic as well as hyperbolic domain walls could form in the absence of externally imposed unidirectional rubbing and the entrapped air bubbles, as an intrinsic feature of the distorted polarization fields $\mathbf{P}(\mathbf{r})$ in samples in which the boundary conditions impose no restriction on the in-plane alignment of $\mathbf{P}$, Fig.5 [76, 77]. The polarization field tends to form vortices to reduce the effects of depolarization field. Consider two situations. In one, the N$_F$ polarization within a sample of an area $A = L^2$ and thickness $h$, is uniform, $\mathbf{P} = \{P_x, P_y, P_z\} = P\{1,0,0\}$. It means that the two $yz$ sides of the sample are charged with the surface densities $\pm P$. The corresponding depolarization field and the electrostatic energy are then $\mathbf{E}_{DP} = -\frac{\mathbf{P}}{\varepsilon\varepsilon_0}$ and $U_{DP} = \frac{P^2 L^2 h}{\varepsilon\varepsilon_0}$, respectively. There is no elastic energy as the polarization is uniform. Consider now a circular disk sample of the same area $A = L^2$ and thickness, in which $\mathbf{P}$ forms a circular vortex; in cylindrical coordinates, $\mathbf{P} = \{P_r, P_\psi, P_z\} = P\{0,1,0\}$. Since $\mathbf{P}$ is everywhere tangential to the surface and since there is no splay, the only energy is that one of the elastic bend [60]: $U_{bend} = \pi K_{33} h \ln \frac{L}{r_{core}} + U_{core}$, where $r_{core}$ and $U_{core} \sim \pi K_{33} h$ are the molecular scale radius and the energy of the vortex' core, respectively; $U_{core}$ brings an inessential contribution to $U_E$ and can be absorbed into the rescaled $r_{core}$. The ratio of the energies of the two structures is then $\frac{U_{DP}}{U_{bend}} = \frac{P^2 L^2}{\pi \varepsilon\varepsilon_0 K_{33} \ln \frac{L}{r_{core}}}$

. With $L = 10$ μm, $\ln\frac{L}{r_{core}} \sim 10$, and the estimates above, $\frac{U_{DP}}{U_{bend}} \sim 10^6$, a huge number. It is only when $L \sim 10$ nm, close to $\xi_P$, that the two states show a similar energy. Of course, the surface polarization charges can be screened by charges of free ions, so that the depolarization field is reduced to $E_{DP} = -\frac{P+\sigma_s}{\varepsilon\varepsilon_0}$. The typical surface charge of adsorbed ions reported for nematics [78, 79] is rather weak, $\sigma_s \sim (10^{-4} - 10^{-5})$ C m$^{-2}$, smaller than $P \approx (4-6) \times 10^{-2}$ C m$^{-2}$. Although higher values of $\sigma_s$ are possible, see the discussion below, one might still expect that vortex states of N$_F$ films with azimuthally degenerate anchoring are energetically similar or even preferrable than extended areas of a uniform polarization.

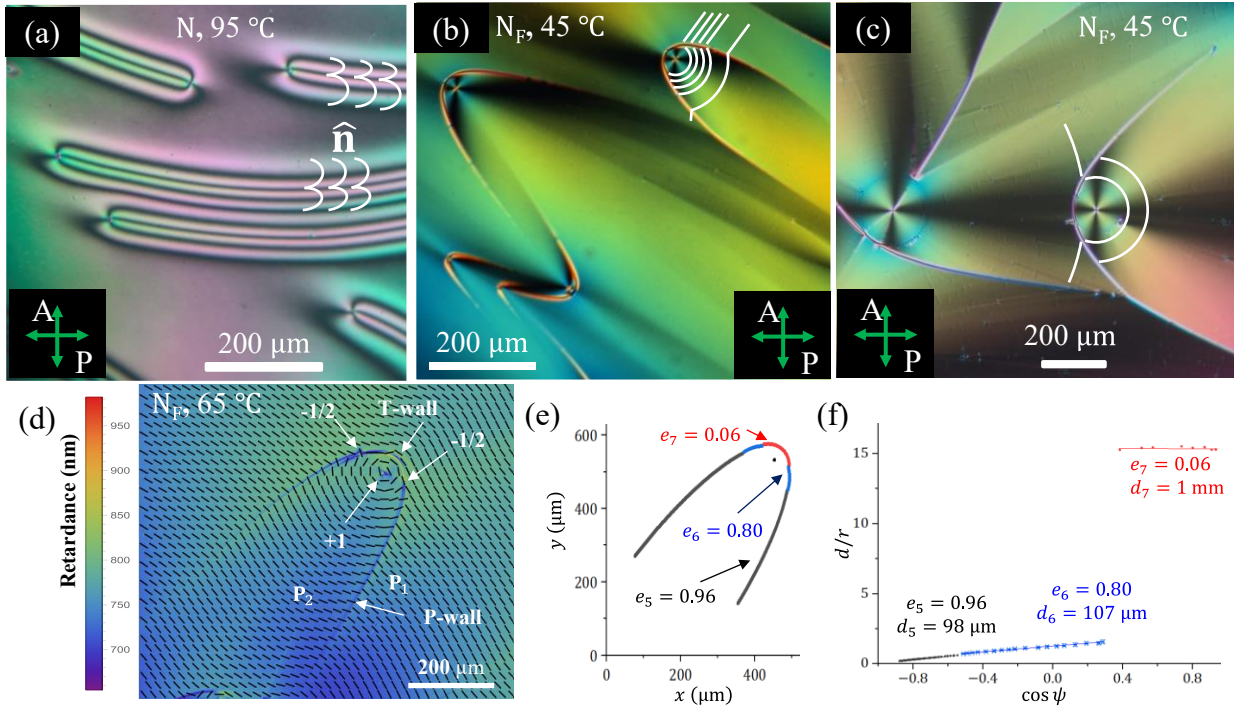

**Figure 5.** DIO N$_F$ thin films at a glycerin substrate. (a) N$_U$ film with $2\pi$ splay-bend domain walls; (b,c), N$_F$ texture of conic-sections with prevailing circular bend and parabolic (b) or (c) hyperbolic domain walls; $\hat{\mathbf{n}}$ is depicted by white lines in (a-c); (d) parabolic domain wall separating a +1 vortex from a relatively uniform domain; note a composite defect representing a T-wall limited by two -1/2 disclinations, visualized by PolScope; (e) eccentricity of different parts of the parabolic wall in (d); (f) fitting the conic in (d) with directrix and eccentricity. Data from Ref.[77].

The samples are prepared by spreading a thin (few micrometers) film of $N_F$ onto the surface of immiscible fluid, such as glycerin. Alternatively, one can use coatings with polymers such as polystyrene that impose no in-plane azimuthal preference for the orientation of $N_U$ [80] and $N_F$ [77]. Since the $N_F$ exhibits no crystallographic axes, these $N_F$ samples set no preferred direction of **P**, except that **P** tends to be tangential to the interface to avoid depositing charges on it.

Figure 5a-c shows the textures of thin DIO films spread onto the surface of glycerin; the upper surface is free. In the $N_U$ phase, hybrid alignment of the director, tangential at the glycerin and homeotropic at the free surface, leads to $2\pi$ domain walls of the "W" director geometry [81]; these $2\pi$ domain walls contain both splay and bend and are clearly distinguished in Fig.5a as bands with four extinction bands. In the $N_F$, the director and **P** are tangential to both the $N_F$-glycerin and $N_F$-air interfaces. The most important feature is that the curvature lines of **P** and $\hat{\mathbf{n}}$ in the $N_F$ are close to circles and circular arches, which implies prevalence of bend and formation of +1 vortices $\mathbf{P} = \{P_r, P_\psi, P_z\} = P\{0,1,0\}$, Fig.5b,c. Each vortex is separated from a uniform or nearly uniform **P** by walls shaped as parts of ellipses and parabolas, Figs. 5b, while two neighboring +1 vortices are separated by hyperbolic walls, Fig. 5c.

The shapes of domain walls are verified with an equation of a conic, written in polar coordinates $(r, \psi)$ centered at the core of a circular vortex, as

$$\frac{d}{r} = \frac{1}{e} - \cos\psi, \tag{2}$$

where $e$ is the eccentricity, $d$ is the distance from the core to the directrix. The domain walls satisfy Eq.(1) with either $e \approx 1$ (parabolic, or P-walls) or $e > 1$ (hyperbolic, or H-walls) everywhere, except for the tip regions. Near the tips, the fits yield a much smaller $e$ characteristic of elliptical and circular arcs; these arcs are abbreviated as T-walls. The T-walls are 180° DWs, separating two antiparallel polarizations, **P** and $-\mathbf{P}$, and limited by two -1/2 disclinations.

The shapes of P- and H-walls are dictated by the $N_F$ tendency to form vortex states and avoid bound electric charge [77]. The equivalent of the bulk bound electric charge is the interfacial charge of density $\sigma_b = (\mathbf{P}_1 - \mathbf{P}_2) \cdot \hat{\mathbf{v}}_1$, where $\hat{\mathbf{v}}_1$ is the unit normal to a domain wall, pointing from domain 2 towards domain 1. Away from the domain walls and the cores of circular vortices, $|\mathbf{P}_1| = |\mathbf{P}_2| = P$. To be uncharged, a domain wall must bisect the angle between $\mathbf{P}_1$ and $\mathbf{P}_2$, so that $\mathbf{P}_1 \cdot$

$\hat{\boldsymbol{v}}_1 = \mathbf{P}_2 \cdot \hat{\boldsymbol{v}}_1$, which means that the components of the polarization along the normal to the wall are continuous and equal each other while the projections onto the wall are antiparallel.

The remarkable bisecting properties of conics, elucidated millennia ago by Apollonius of Perga [82], are often formulated in terms of light reflection [83]. Consider a parabola, Fig.6a. Light emitted from a focus, which is the core of the circular vortex in the $N_F$ case, is reflected by the parabola along the lines parallel to the symmetry axis. A tangent to a parabola at a point $(x, y)$ makes equal angles with the radius-vector directed from the focus and with the reflected beam. Equivalently, the angle $\theta_1$ between $\mathbf{P}_1$ and the P-wall and the angle $\theta_2$ between $\mathbf{P}_2$ and the P-wall are equal, Fig.6a,

$$\theta_1 = \theta_2 = \arctan\sqrt{\frac{x}{f}}, \tag{3}$$

where $\eta =$ and the origin of the Cartesian coordinates $(x, y)$ is at the conic's vertex. Therefore, when a P-wall separates a circular vortex of $\mathbf{P}_2$ from a uniform domain with $\mathbf{P}_1$ orthogonal to the parabola's axis, its parabolic shape guarantees that $\mathbf{P}_1 \cdot \hat{\boldsymbol{v}}_1 = \mathbf{P}_2 \cdot \hat{\boldsymbol{v}}_1$ and carries no surface charge, $\sigma_b = 0$. The bulk charge $\rho_b$ is also zero since there is no splay of $\mathbf{P}_1$ and $\mathbf{P}_2$. The H-wall features a similar bisecting property which assures a zero $\sigma_b$ at the boundary between two vortices, Fig.6b.

One expects that the variation of the projection of $\mathbf{P}$ onto $\hat{\boldsymbol{v}}_1$ across the wall would create two sheets of opposite charges if $\mathbf{P}$ remains in the plane of the sample [70, 72, 75], as in a Néel wall in ferroelectric crystals; these two charged sheets are shown in Fig.4. Note that in the textures in Fig.5, the width $w$ of the P- and H- walls is $\sim 10$ μm [77], much wider than $\xi_P \sim \sqrt{\frac{\varepsilon \varepsilon_0 K}{P^2}} \sim 1$ nm. One reason is that the polarization is screened by ions, so that the domain width can be estimated as $w \sim \sqrt{\frac{\varepsilon \varepsilon_0 K}{(P+\sigma)^2}}$. With $\varepsilon \varepsilon_0 \sim 10^{-9} \frac{C^2}{J \times m}$, $K \sim 10^{-11}$ N, $w \sim 10$ μm, one finds $P + \sigma \approx 10^{-5} \frac{C}{m^2}$, a three orders of magnitude reduction from the un-screen polarization. One should not exclude also the possibility of the twist of $\mathbf{P}$, either around an in-plane axis, as in a Bloch wall, or around the twist axis which is perpendicular to the film.

The combined defects representing a T-wall sandwiched between two -1/2 disclinations are caused by the fact that the bend angle $\delta$ between the vectors $\mathbf{P}_1$ and $\mathbf{P}_2$ near the vertex increases, making the polarization field "hairpin"-like, Fig.6c. The -1/2 disclinations replace this large bent angle $\delta = \pi - 2\theta$ with two small angles $\beta \approx \theta$, thus reducing the bend energy [77].

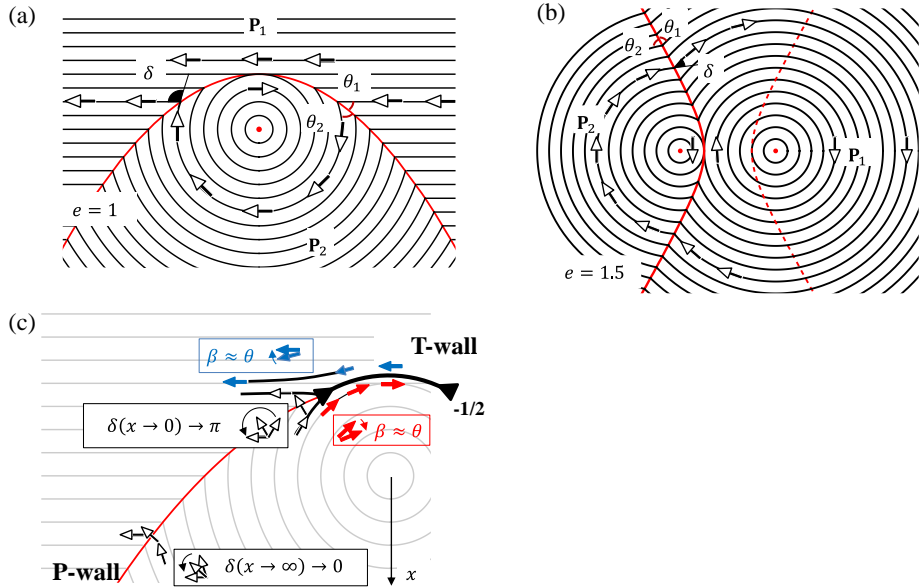

**Fig.6.** Schemes of (a) parabolic, (b) hyperbolic, and (c) T-walls sandwiched between two -1/2 disclinations. Data from Ref.[77].

The composite defects representing T-walls bounded by half-integer disclinations have been predicted for $N_F$ [84, 85] as analogs of the domain walls seeded by cosmic strings in the early Universe models [86] and of domain walls bounded by half-quantum vortices recently found in a superfluid $^3$He [87, 88]. In the Universe and $^3$He scenarios, the composite domain walls appear after a phase transition from a symmetric phase that contains isolated strings/disclinations. In the less symmetric phase, the isolated disclinations are topologically prohibited and must be connected by a domain wall. In contrast, the -1/2 disclinations at the ends of T-walls described by Kumari et al. [77] serve to reduce the elastic energy of strong bends, Fig.6c, and appear without any reference to the potential seeds in the more symmetric phase.

In 1910, G. Friedel and F. Grandjean [89] described ellipses and hyperbolas seen under a microscope in a liquid crystal of a type unknown at that time. A later analysis [90, 91] revealed that these conics are caused by a layered structure of the liquid crystal known nowadays as a smectic A (SmA). The layers are flexible but preserve equidistance when curled in space. The director $\hat{\mathbf{n}}$ is normal to the equidistant layers and can experience only splay but not twist nor bend. The families of flexible equidistant surfaces form focal surfaces at which the layers curvatures diverge. To reduce the energy of these singular focal surfaces, the SmA reduces them to lines of confocal conics, such as an ellipse-hyperbola or two parabolas [92]; these pairs form the frame of the celebrated focal conic domains (FCDs) [60]. Gray lines in Fig. 6 could be interpreted as cuts of smectic layers wrapped around a parabola and hyperbola of FCDs, Fig.7. The $N_F$ conics are shaped by a different mechanism, rooted in the avoidance of the space charge. In the $N_F$, $\hat{\mathbf{n}}(\mathbf{r})$ and $\mathbf{P}(\mathbf{r})$ tend to be solenoidal, div$\hat{\mathbf{n}}$ = div$\mathbf{P}$ = 0, while the director in SmA is irrotational, curl$\hat{\mathbf{n}}$ = 0. Besides this difference in physical underpinnings, there is also a distinction in how the conics in the $N_F$ and SmA heal cusp-like singularities. In the $N_F$, the cusps are attended by a bend of the polar vector $\mathbf{P}$, which necessitates the -1/2 disclinations and the T-walls at the tips of the conics, while in the SmA, a similar cusp could be healed by weak splay of the apolar director $\hat{\mathbf{n}} \equiv -\hat{\mathbf{n}}$.

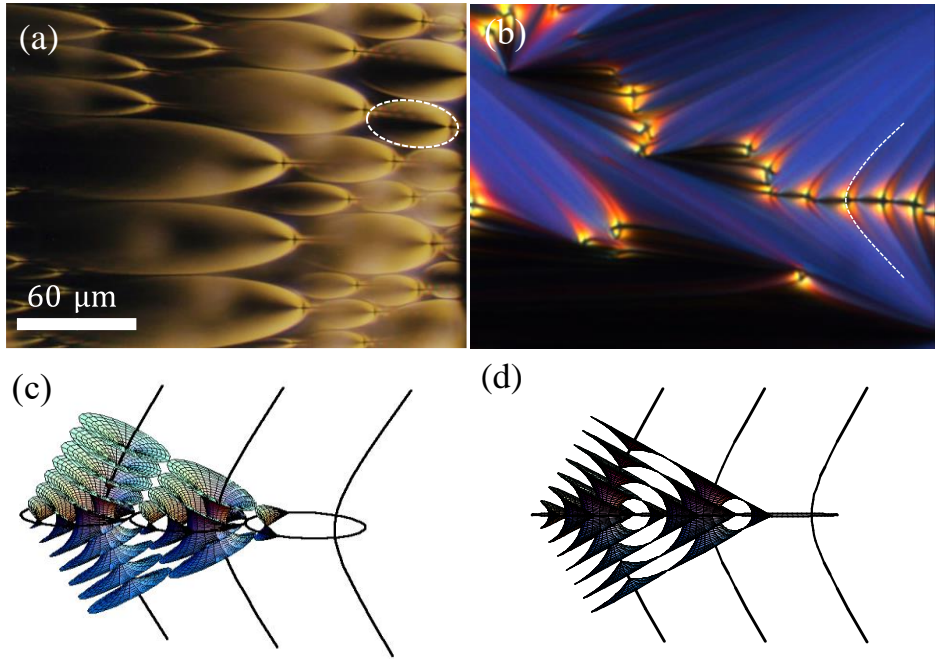

**Fig. 7.** Smectic A textures of focal conic domains with (a) ellipses in the plane of the sample; (b) hyperbolas in the plane of the sample; (c,d) two 3D views of SmA domain walls formed by focal conic domains.

**CONCLUSION**

As indicated by de Gennes [1], the full form of the Frank-Oseen energy (1) is "too complex to be of practical use," either because the values of the corresponding elastic constants are not known or because the equations are prohibitively difficult to solve. One often resorts to the so-called one-constant approximation in which all constants are assumed to be equal. The presented examples underscore the importance of elastic constants disparity. The problems might still be "simple" when only a few types of distortions are at play, such as splay and saddle-splay in the description of FCDs in SmA [93]. So far, the textures of the $N_F$ have been explored for relatively thin quasi-2D films. Bulk samples with a 3D divergence-free polarization field might reveal more complex structures. For example, I. Luk'yanchuk et al. [94] predicted that a small spherical particle of solid ferroelectrics should produce a hopfion, which is a set of interlinked circles. Hopfions have been already observed in liquid crystals such as nematic-based ferromagnets by I.I. Smalyukh et al. [95, 96, 97]. A need for a hopfion in 3D can be justified by the argument that **P** is everywhere tangential to the spherical surface but instead of forming a singular vortex-like disclination, it features a core with **P** escaped into the third dimension [94], a notion well known in the physics of disclinations in liquid crystals [98, 99]. Droplets of $N_F$ might be a natural home for hopfions.

The list of new discoveries has been recently extended by the twist-bend ferroelectric nematic $N_{TBF}$, synthesized at the Military University of Technology in Poland [100]. The new phase, formed by achiral polar molecules, with a spontaneous electric polarization along the heliconical axis, is a ferroelectric analog of the paraelectric $N_{TB}$. The pseudolayers of the $N_{TBF}$, associated with the constant phase of the molecular tilts, tend to keep equidistance, which hinders the twist and bend of the heliconical axis. On the other hand, splay of this axis is also hindered, since it creates space charge. Remarkably, the pitch of heliconical $N_{TBF}$ structure is in the submicron range and changes under an externally applied dc electric field. At higher field, the structure shows a shorter pitch and a smaller conical angle, eventually unwinding into a uniform nematic structure, a behavior analogous to the paraelectric response of the oblique helicoidal cholesteric [40, 41, 101].

Kim, R. Koizumi, P. Kumari, B-X. Li, M. Rajabi, J. Xiang, who performed most of the experiments reviewed in this paper, and to the organizers of the 16$^{th}$ European conference on liquid crystals at the University of Calabria in Rende, Italy, for the opportunity to present these results.

**Disclosure statement**

No potential conflict of interest was reported by the author.

**Funding**

The work is supported by the NSF grants DMS-2106675, DMR-2122399, and DMR-2215191.

**ORCID**

Oleg D. Lavrentovich http://orcid.org/0000-0002-0128- 0708

**Author contributions**

O.D.L. wrote the paper.

**Data availability statement**
The data that support the findings of this study are available within the manuscript.